\documentclass[12pt]{iopart}

\usepackage{graphicx}% Include figure files
\usepackage{bm}% bold math
\usepackage{hyperref}% add hypertext capabilities
\usepackage{marginnote}
\renewcommand{\marginnote}[1]{} % disable marginnotes temporarily

\usepackage{xcolor} 
\usepackage{ulem}
\usepackage{siunitx}
\usepackage{upgreek}
\usepackage{enumerate}

\newcommand{\YRS}{YbRh$_{2}$Si$_{2}$}

\begin{document}

\title[Microstructuring \YRS]{Microstructuring \YRS\ for resistance and noise measurements down to ultra-low temperatures}

\author{Alexander Steppke$^{1,2,3}$, Sandra Hamann$^{1,4}$, Markus K\"onig$^1$, Andrew P. Mackenzie$^{1,5}$, Kristin Kliemt$^6$, Cornelius Krellner$^6$, Marvin Kopp$^6$, Martin Lonsky$^6$, Jens M\"uller$^6$, Lev V.\ Levitin$^7$, John Saunders$^7$, Manuel Brando$^1$
}%

\address{$^1$ Max-Planck-Institute for Chemical Physics of Solids, Noethnitzer Str. 40, 01187 Dresden, Germany}
\address{$^2$University of Zurich, Winterthurerstr. 190, 8057 Zurich, Switzerland}
\address{$^3$Paul Scherrer Institute, Forschungsstr. 111, 5232 Villigen, Switzerland}
\address{$^4$Helmholtz-Zentrum Dresden-Rossendorf, Bautzner Landstr. 400, 01328 Dresden, Germany}
\address{$^5$Scottish Universities Physics Alliance, School of Physics and Astronomy, University of St Andrews, St Andrews, UK}
\address{$^6$Physikalisches Institut, Goethe-Universit\"at Frankfurt, Max-von-Laue-Str. 1, 60438 Frankfurt am Main, Germany}
\address{$^7$Department of Physics, Royal Holloway University of London, Egham, Surrey TW20 0EX, UK}
\date{\today}
\begin{abstract}
The discovery of superconductivity in the quantum critical Kondo-lattice system \YRS\ at an extremely low temperature of 2\,mK has inspired efforts to perform high-resolution electrical resistivity measurements down to this temperature range in highly conductive materials.
Here we show that control over the sample geometry by microstructuring using focused-ion-beam (FIB) techniques allows to reach ultra-low temperatures and increase signal-to-noise ratios (SNR) tenfold, without adverse effects to sample quality.
In five experiments we show four-terminal sensing resistance and magnetoresistance measurements which exhibit sharp phase transitions at the N\'eel temperature, and Shubnikov-de-Haas (SdH) oscillations between $13\,$T and $18\,$T where we identified a new SdH frequency of 0.39\,kT. The increased SNR allowed resistance fluctuation (noise) spectroscopy that would not be possible for bulk crystals, and confirmed intrinsic $1/f$-type fluctuations. Under controlled strain, two thin microstructured samples exhibited a large increase of $T_N$ from 67\,mK up to 188\,mK while still showing clear signatures of the phase transition and SdH oscillations. SQUID-based thermal noise spectroscopy measurements in a nuclear demagnetisation refrigerator down to 0.95\,mK, show a sharp superconducting transition at $T_c = 1.2$\,mK. These experiments demonstrate microstructuring as a powerful tool to investigate the resistance and the noise spectrum of highly conductive correlated metals over wide temperature ranges.
\end{abstract}


\maketitle

\section{Introduction}
\marginnote{R(T) motivation}
A temperature-dependent measurement of the resistivity of a material provides several insights into its physical properties. For instance, we can observe subtle changes in the electronic structure such as spin or charge density waves~\cite{Kummamuru2008}, structural phase transitions~\cite{Bartlett2021} and more dramatic changes such as superconductivity.

\marginnote{Lock-in amplifiers}
The development of lock-in amplifiers in the early 1950s~\cite{Stutt1949} has simplified resistance measurements by extracting signals well below the noise floor~\cite{PerkinElmer2000}, so that nowadays a measurement of the resistivity of very good metals has become standard practice~\cite{Temple1975}. 
\marginnote{Thermometry example}
Moreover, the temperature-dependent resistivity of particular materials allows their use as high-precision secondary thermometers~\cite{Pobell2007}, often over wide temperature ranges.

\marginnote{R(T) difficult at low T}
High-resolution resistance measurements can easily be performed under ambient conditions. However, at very low temperatures (milli- or microkelvins), a key limitation of standard four-terminal sensing techniques, with either voltage or current excitation, is the dissipation, which must be minimized. Otherwise, insufficient thermalization leads to temperature gradients or generally prevents cooling to lowest temperatures.
Dissipation in the form of Joule heating is mainly caused by the electrical resistance of the contacts to the sample $R_\mathrm{c}$ and the sample's own resistance $R_\mathrm{s}$, when driven by an either constant- or time varying excitation current $I_\mathrm{exc}$.

We therefore need to find a balance in any low temperature resistance measurement between a sufficiently large measured sample voltage $U_{\mathrm{s}}$ and minimizing the dissipated power $P_\mathrm{diss}$, which are given by
\begin{equation}
U_{\mathrm{s}} = R_\mathrm{s} \cdot I_\mathrm{exc}
\end{equation}
and
\begin{equation}
P_\mathrm{diss} = \left(R_\mathrm{s} + R_\mathrm{c}\right) \cdot  I_\mathrm{exc}^2. \label{eq:dissipation}
\end{equation}
In many classes of correlated materials, e.g. oxides such as the cuprate superconductors, $R_\mathrm{c}$ can be higher than several ohms and, even for metallic samples, can reach hundreds of milliohms, exceeding $R_\mathrm{s}$ and dominating the dissipation.

A straightforward approach to this optimization problem is to reduce $I_\mathrm{exc}$, thereby exploiting the quadratic dependence in the dissipated power. Addressing the corresponding reduction in signal amplitude led to the development of a wide range of techniques. These range from electrical circuits using a bridge-type setup, which were crucial for the discovery of superconductivity~\cite{Onnes}, over the aforementioned lock-in amplifiers to amplification techniques involving low-temperature transformers~\cite{Haselwimmer2001} or advanced superconducting quantum interference device (SQUID) amplifiers~\cite{Drung2007}.

Still, there is a limit to this approach, especially when measuring metallic samples with high conductivity. The reduction in signal amplitude by reducing the excitation current produces a setup which is susceptible to any external electrical disturbances. In addition, the smaller signal leads to a decrease in resolution, which can make the extraction of features such as quantum oscillations or topological transitions infeasible. 

With modern microstructuring techniques, a new and different path opens up to address this trade-off between dissipation and achievable resolution. With precision control over the sample geometry, one can increase the ohmic resistance $R_{\mathrm{s}}$ by creating thin bar-shaped or meander-like samples. We will show that it is possible to increase $R_{\mathrm{s}}$ by a factor 250 (up to some $\Omega$) while keeping $R_{\mathrm{c}}$ small (a few m$\Omega$). We can then distinguish two regimes in the measured signal amplitude 
\begin{equation}
	U_\mathrm{signal} = \sqrt{\frac{R_\mathrm{s}^2 \, P_\mathrm{diss} }{R_\mathrm{s} + R_\mathrm{c}}} \ 
\end{equation}
for a given fixed maximum dissipation. For $R_\mathrm{s} \ll R_\mathrm{c}$, an increase in $R_\mathrm{s}$ is directly proportional to the amplitude and for higher $R_\mathrm{s}$, we observe a square-root increase of the amplitude.
The assumption of a fixed maximum dissipation is often fulfilled when the sample is thermally strongly linked to its environment. Here lies also the limitation of this approach. For extreme changes in the geometry, one can reduce the thermal coupling between sample and cold bath significantly. Under these semi-adiabatic conditions, the decrease in thermal conductance to the cold bath outweighs the reduced dissipation. 

From these considerations we can devise an optimum geometry starting from a irregularly shaped bulk sample with approximate cross section area $A$ and length $l$. The heat capacity of such a sample is then given by $C = c \,\rho_m\, A\, l$ with the density $\rho_m$ and the specific heat $c$. If possible, we do not want to decrease the  heat capacity but only increase the resistance given by $R_\mathrm{s} = \rho\, l / A$. This can be achieved by cutting thin slots, creating a meander-like structure as shown in figure \ref{fig:YRS_fig1_rho_T}(a). Even by cutting only a single narrow slot ($A\rightarrow A/2$, $l \rightarrow 2 l$) into a rectangular sample, one can leave $C$ almost unchanged while increasing the resistance fourfold. 

Whereas these considerations were focused on a resistance measurement using a four-terminal technique, alternative approaches can also benefit from the same structuring technique. For measurements at lowest possible temperatures, a further decrease in dissipation can be achieved by removing the drive current completely and deducing the resistance from the sample's Johnson noise. Due to the small size of this effect ($\ll 10^{-9}$\,V at 10\,mK), very high amplification is required, which can be achieved by inductive coupling to a SQUID~\cite{Johnson1928,Giffard1972,Webb1973}. Although this technique has been optimized by using modern SQUIDs~\cite{Drung2007} and new circuit designs~\cite{Casey2014}, so that the resistance of the sample itself can be of the order of $0.1-1$\,m$\Omega$, inherently the circuit loop includes the contact resistances in series with the sample's own resistance. In practice, for bonded or spot-welded contacts with small wires (25-50\,$\upmu$m in diameter), this can also be of the order of 0.1 to 1\,m$\Omega$, but in many cases still much larger than that of samples from highly conductive materials. Therefore, a higher sample resistance can be beneficial for a wide range of techniques, and we outline several ways this can be achieved by changing the geometry using microstructuring.

In single crystals of highly conductive materials with $\rho_{0} \leq 1$\,$\upmu\Omega$cm, an increase of the resistance is difficult to achieve with standard sample preparation techniques, even for a modest resistance of $R_\mathrm{s} = 0.1$\,$\Omega$. We would need to cut a sample so that the length to cross-sectional-area ratio $l/A$ reaches $10^{5}$\,cm$^{-1}$ to achieve a resistance comparable to the contact resistance. This means that even if the growth of large crystals with $l\approx$1\,cm is feasible, the cross-sectional-area needs to be of the order of $1000$\,$\upmu$ m$^{2}$. Assuming that the sample's width could be cut to $100\,\upmu$m, its thickness has to be reduced to $10$\,$\upmu$m, which is difficult to achieve and even more difficult to handle manually. These established sample preparation techniques, e.g. using diamond wire saws, polishing and lapping, are therefore not suitable, and more complex geometries like a meander-shaped resistance track are required to achieve significant resistances increases.

\marginnote{FIB to the rescue}
In contrast to the aforementioned methods and other microstructuring techniques (see Appendix), the recent development and commercial availability of focused-ion-beam (FIB) devices (see also Appendix) allows such a sample manipulation. These devices can shape and contact samples on sub-micrometer dimensions without altering their physical properties~\cite{Moll2010}. Thin metallic platelets can be cut in long patterned meander tracks of width of less than 5\,$\upmu$m~\cite{Moll2016} and along user-selected crystalline directions \cite{Putzke2020}. With a track width of 5\,$\upmu$m and a typical thickness of 20\,$\upmu$m for thin samples, $A =  10^{-6}$\,cm$^{2}$, which in the corresponding example given above would result in $R = 1$\,$\Omega$. 

Such a strong increase in resistance allows resistivity measurements at much lower temperatures and higher signal-to-noise ratio (SNR), and therefore to access a wide range of phenomena such as superconductivity below 10\,mK and voltage noise spectra at classical and quantum phase transitions, and to resolve subtle signatures, for example those caused by topological transitions \cite{Pfau2017}.

Here we demonstrate this approach for the strongly correlated electron system \YRS~\cite{Trovarelli2000,Custers2003,Gegenwart,Gegenwart2008}, a canonical heavy-fermion (HF) compound. One of the main motivations for this work was the discovery of unconventional superconductivity (SC) in \YRS~\cite{Trovarelli2000,Schuberth2016} at the very low critical temperature $T_{c} = 2$\,mK~\cite{Schuberth2016}. Whereas this was demonstrated by ac-susceptibility measurements~\cite{Schuberth2016,Steinke2017}, a resistivity measurement was deemed infeasible in as-grown crystals, as the dissipation is prohibitively high in a four-terminal measurement. Only recently, transport experiments in highly optimized setups \cite{Nguyen2021,Levitin2022} corroborated the initial findings.

The discovery of superconductivity in \YRS\ was unexpected because it is a rare phenomenon in Yb-based HF compounds, with $\beta-$YbAlB$_{4}$ ($T_{c} = 80$\,mK)~\cite{Nakatsuji2008} the only other known superconducting material in this class. On the other hand, there are many Ce-based (Ce is considered the electron analog of Yb) HF superconductors with critical temperatures as high as 2.3\,K~\cite{{Petrovic2001}}. 
Whereas the particularities of the superconducting state in \YRS\ are still under investigation, it has been suggested that the suppression of antiferromagnetic (AFM) order ($T_{N} = 70$\,mK~\cite{Trovarelli2000}) by nuclear ordering promotes the emergence of superconductivity~\cite{Schuberth2016,Steinke2017}. Such a concomitant suppression of magnetic order with the appearance of superconductivity spans wide ranges of material families beyond other HF systems~\cite{Steglich1979,Mathur1998}, to cuprates~\cite{Lee2006}, pnictides  or organic charge-transfer salts~\cite{Toyota2007,Kanoda2008}. If the mechanism behind HF superconductivity is connected to vanishing magnetic order, we can speculate that in Yb-based materials with their lower ordering temperatures, SC also appears at much lower temperatures (in the mK range) compared to Ce-based compounds. A reliable way of measuring resistance at those temperatures will therefore facilitate the discovery of more Yb-based superconductors. In addition, this enables measuring the intrinsic noise spectrum of a high-resistance sample of a material close to a classical or quantum phase transition, which can provide additional information about the critical fluctuations.

In the following we illustrate microstructuring approaches and their advantages in \YRS, using electrical resistivity and noise measurements in five experiments from three meander-like and two bar-like samples:
\begin{enumerate}[i]
    \item Study of the resistivity exponent across $T_{N}$ on meander \#1
    \item Magnetoresistance and Shubnikov-de-Haas oscillations on meander \#2
    \item High-temperature noise spectroscopy on meander \#3
    \item Tuning of $T_N$ by strain on meander \#2, bar \#1 and bar \#1 after reprocessing
    \item Ultra-low temperature noise measurements on meander \#2
\end{enumerate}

\section{Study of the resistivity exponent across $T_{N}$ on meander \#1}
\begin{figure}[htb]
	\centering
	\includegraphics[width=\linewidth]{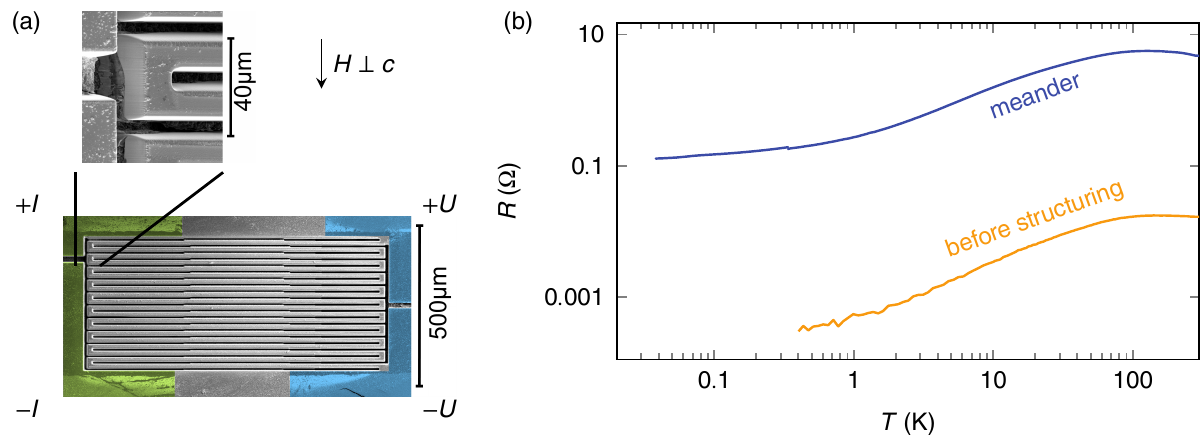}
	\caption{(a) Electron microscope image of the \YRS\ meander \#1. The four-terminal contacts for electrical measurements are shown in green (current) and blue (voltage). The meander is structured from the ab-plane of the crystal, the applied field direction is perpendicular to the long sides of the meander and perpendicular to the c-axis.
		(b) Temperature dependence of the resistance of the \YRS\ sample before and after FIB structuring. The electrical contacts using bonded gold wires remained on the sample during microstructuring.}
	\label{fig:YRS_fig1_rho_T}
\end{figure}

For our first experiment, a flat, unpolished \YRS\ crystal~\cite{Krellner} with a size of approx. $1.8\,\text{\,mm} \times 1.2\,\text{mm}$  %$L \times W \approx 1.8\,\text{\,mm} \times 1.2\,\text{mm}$ 
was fixed onto a Si substrate using Araldite Rapid. We then used a shadow-mask technique and standard electron-beam evaporation of Au to deposit contacts onto the \YRS\  crystal and connect the crystal to Au leads and contact pads on the Si substrate. The FIB patterning of the crystal into the meander was done in a standard FIB/SEM system (FEI Helios Nanolab G3), using an acceleration voltage and Ga-ion beam current of 30\,kV and 65\,nA, respectively, for the initial steps. Lower values were used in subsequent steps to minimize the damage to the crystal. The meander structure, shown in figure~\ref{fig:YRS_fig1_rho_T}(a), consists of 20 straight wire-like segments with dimensions of $L \times W \approx 900\,\upmu\text{m} \times 14\,\upmu\text{m}$ each, resulting in a total wire length of 18.8\,mm including the bends connecting the straight segments, i.e. $l/W \approx 1250$. Due to the necessary large size of the initial bulk crystal the meander thickness varies between $15\,\upmu\text{m}$ and $75\,\upmu\text{m}$ (cf. Tab.~\ref{tab.Samples_strain.effect}).

\marginnote{PPMS and Kelvinox description}
Four-terminal resistivity measurements were performed using a Quantum Design PPMS from room temperature to 0.4\,K and using an Oxford Instruments dilution refrigerator down to 20\,mK equipped with low-temperature transformers~\cite{Haselwimmer2001}.

\begin{figure}[htb]
	\centering
	\includegraphics[width=\linewidth]{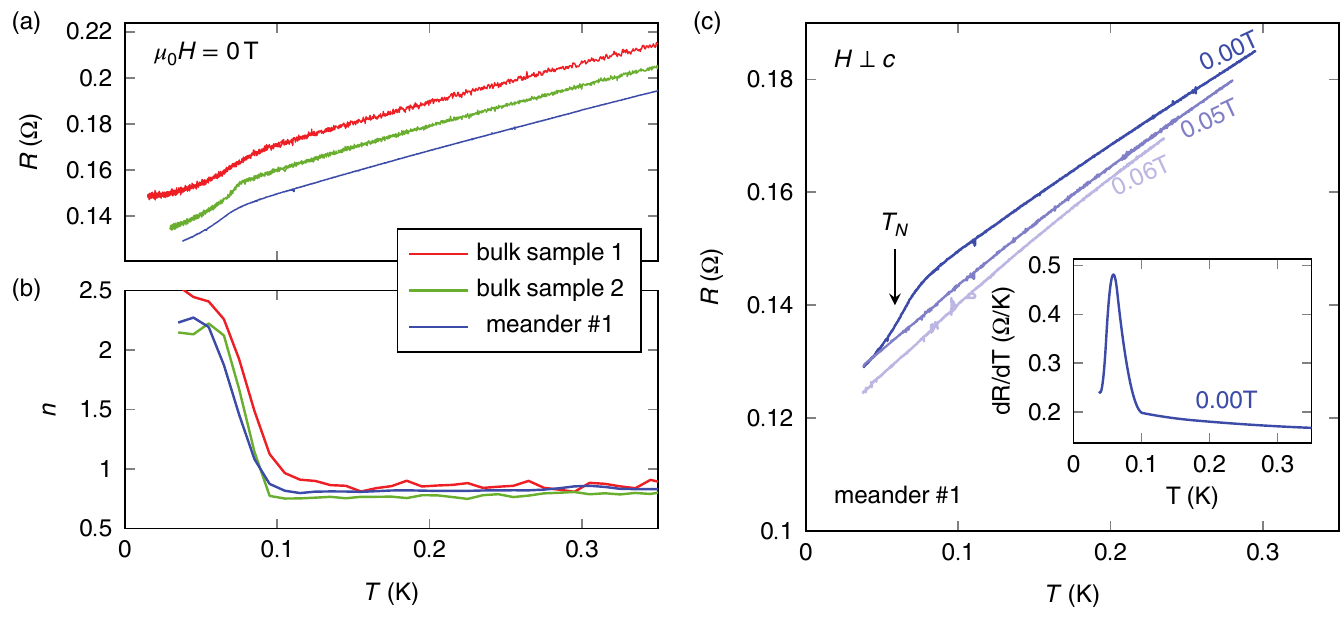}
	\caption{(a) Low-temperature electrical resistance $R$ of the microstructured meander \#1 at zero field. For comparison the electrical resistance of two bulk samples (sample 1 from Ref.~\cite{Steglich2014} and sample 2 from Ref.~\cite{Gegenwart2008}) are scaled on the meander \#1 data and shifted subsequently with an offset of $+0.01\,\Omega$ for clarity.
	 (b) Resistance exponent $n$ from the analysis with $R(T) = R_0 + A\cdot T^{n}$ above and within the AFM ordered phase.
	(c) Low temperature electrical resistance $R$ of the microstructured meander \#1 measured at different magnetic fields applied along the basal plane ($B \perp c$). Inset: Temperature derivative of the electrical resistance at zero field. The peak indicates $T_{N} = 67$\,mK, close to values reported for previous sample growths of $70$\,mK \cite{Gegenwart2002} and $72$\,mK \cite{Brando2013}.
    } 
\label{fig:YRS_fig2_rho_n_H}
\end{figure}

Resistance measurements in the temperature range from 300$\,$K to 30$\,$mK before and after microstructuring are shown in Figure~\ref{fig:YRS_fig1_rho_T}. Microstructuring led to an increase of the resistance by a factor of 250. Although the thickness of the sample is not uniform, we can estimate a geometry factor of $l/A = l/ (d_\mathrm{mean} \cdot W)= 18,800\upmu \text{m}/(45 \upmu \text{m} \cdot 14\upmu \text{m}) = 630 \upmu \text{m}^{-1}$.
Both measurements can nearly be scaled on top of each other, although we observed some undesired parallel conductance path through the Si substrate near room temperature. They show the expected Kondo-lattice behavior with a coherence maximum at $T^{*} \approx 100$\,K~\cite{Trovarelli2000} and a strong drop for $T < T^{*}$ followed by the linear-in-$T$ dependence of the resistivity at low temperatures (see figure~\ref{fig:YRS_fig2_rho_n_H}(c)), characteristic of the quantum critical behavior of this material~\cite{Custers2003}.

We focus now on the low-temperature properties of meander \#1 as shown in figure~\ref{fig:YRS_fig2_rho_n_H}: A kink in the resistance is evidence of the onset of AFM ordering (figure~\ref{fig:YRS_fig2_rho_n_H}(a),(c)). Comparable to former measurements on bulk samples $T_N$ is approximately 67\,mK~\cite{Trovarelli2000,Custers2003}. Applying a magnetic field of $B_{N} = 0.06$\,T along the basal plane suppresses the AFM ordering completely (see figure~\ref{fig:YRS_fig2_rho_n_H}(c))~\cite{Gegenwart2002,Brando2013}.  

To visualize the improvement in the obtained SNR we compare measurements on our microstructured meander \#1 with those on two bulk samples taken with higher excitation currents (cf. figure~\ref{fig:YRS_fig2_rho_n_H} and Tab.~\ref{tab:sample_noise_properties}). Here the measured resistance fluctuations are approximately normally distributed and we define the $\text{SNR}$ as $ R_{T \rightarrow 0}/R_{rms}$, with $R_{T \rightarrow 0}$ the lowest measured resistance and $R_{rms}$ being the root-mean square value of the deviation from a noise free measurement in the temperature range up to $0.8\,$K.

Bulk sample 1 (from Ref.~\cite{Steglich2014}) is, with an RRR of 71, comparable to the sample we used in this study, whereas bulk sample 2 (from Ref.~\cite{Gegenwart2008}) is of higher quality with RRR = 136. Because the resistivity of all three samples is linear-in-$T$ for $T > T_{N}$ and with nearly identical slope, we can scale all resistances onto that of the structured meander \#1 (figure~\ref{fig:YRS_fig2_rho_n_H}a).

It can be immediately seen that the data for the meander sample exhibits a significantly lower noise level. Whereas at $T = 0.15$\,K for bulk samples 1 and 2 the observed signal to noise ratio $R_{rms}/R_0$ is $0.45\%$ and $0.48\%$, respectively, it decreases by a factor of 10 to $R_{rms}/R_0=0.047\%$ in the meander \#1. Although, in comparison with bulk sample \#2 a 50 times smaller excitation current was used in the meander structure, the SNR increases by one order of magnitude.

\begin{table}[htbp]
\caption{\label{tab:sample_noise_properties} Properties of samples used in figure~\ref{fig:YRS_fig2_rho_n_H}: The RRR is defined as $\rho_{T=300\,\mathrm{K}} / \rho_{0}$ with $\rho_{T=300\,\mathrm{K}} = 75$\,$\upmu\Omega$cm~\cite{Custers2004,Westerkamp2009}. Resistance noise is obtained from  voltage noise and taking into account the amplification of the low temperature transformer (1:100) used in all measurements.
}
  \begin{tabular}{@{}llll}
  \br
		Sample  &   meander \#1     & Bulk 1 (\cite{Steglich2014}) & Bulk 2 (\cite{Gegenwart2008}) \\
  \mr
		RRR & 35 & 71 & 136 \\
		Equivalent voltage noise at 0.15K (V/$\sqrt{\text{Hz}}$)       & $1.1\cdot 10^{-12}$ & $8.8\cdot 10^{-14}$ & $3.3\cdot 10^{-14}$ \\
		Resistance noise at 0.15K ($\Omega$/$\sqrt{\text{Hz}}$)       & $1.1\cdot 10^{-8}$ & $1.8\cdot 10^{-10}$ & $6.6\cdot 10^{-12}$ \\
		$R_\text{rms}/R_0$         & 0.047\%               & 0.45\%               & 0.48\%               \\
        SNR         & 2142               & 222               & 208               \\		
		Excitation current ($\upmu$A)                    & 1              & 5                & 50            \\
		$\rho_0$ ($\upmu \Omega$cm)             & --   & 1.05                  & 0.55 \\   

  \br

	\end{tabular}
\end{table}

The decrease in the observed relative resistance fluctuations $R_\mathrm{rms}/R_0$ is not only useful in itself to detect subtle features in the $R(T)$ profile, but it can additionally be used in a more precise analysis of derived quantities. For instance, an important quantity in systems close to a QCP is the temperature dependence of the resistivity which provides information about the nature of the electron scattering~\cite{Rosch1999,Custers2003}. Typical resistance $R(T)$ curves are analyzed according to the function $R(T) = R_{0} + AT^{n}$, where $R_{0}$ is the residual resistance, $A$ is a measure of the electron–electron scattering rate and $n$ is the exponent that in the standard theory of metals - the Fermi-liquid theory - is expected to be equal to 2~\cite{Nozieres2018}. The anomalous exponent $n \approx 1$ observed in \YRS\ is strong evidence for the presence of quantum critical fluctuations and the existence of a new class of an electron fluid~\cite{Custers2003}. Therefore, such an analysis would strongly benefit from high-resolution raw data.

In figure~\ref{fig:YRS_fig2_rho_n_H}(b) we show the temperature evolution of $n(T)$ obtained from a least-squares fit using the aforementioned $R(T)$ using a sliding window of $0.04\,$K. Here $R_0$ is determined from the temperature dependence below $T_N$ and then kept fixed for higher $T$. In comparison between the bulk sample 1 and the microstructured sample we observe a decrease in the relative uncertainty of $\Delta n/n$ above \SI{100}{\milli\kelvin} from $0.05$ to $0.01$.

\section{Magnetoresistance and Shubnikov-de-Haas oscillations on meander \#2}

\marginnote{figure4 Lifshitz transitions}
When applying magnetic fields beyond the suppression of the AFM order, i.e. $B > B_{N}$, several topological Lifshitz transitions have been reported in \YRS\ in the field range up to 20\,T~\cite{Rourke2008,Zwicknagl2011,Pfau2013,Pourret2013} with associated changes of the Fermi surface~\cite{Rourke2008,Westerkamp2009,Sutton2010,Pourret2019}. Because of the low quality of the samples and the limited resolution of the measurements, only three Lifshitz transitions were first observed using combined thermopower and resistivity measurements~\cite{Pfau2013,Naren2013}. A more precise study on better crystals revealed the presence of 8 Lifshitz transitions for fields up to 13\,T applied along the [110] direction~\cite{Pourret2013,Pourret2019}. Furthermore, de Haas–Van-Alphen (dHvA) and Shubnikov–de-Haas (SdH) studies were able to detect parts of the predicted Fermi surface structure, but the limited number of observed frequencies do only partially correspond to theoretical predictions~\cite{Knebel2006,Rourke2008,Westerkamp2009,Sutton2010,Friedemann2013}.

\begin{figure}[t]
	\centering
	\includegraphics[width=\linewidth]{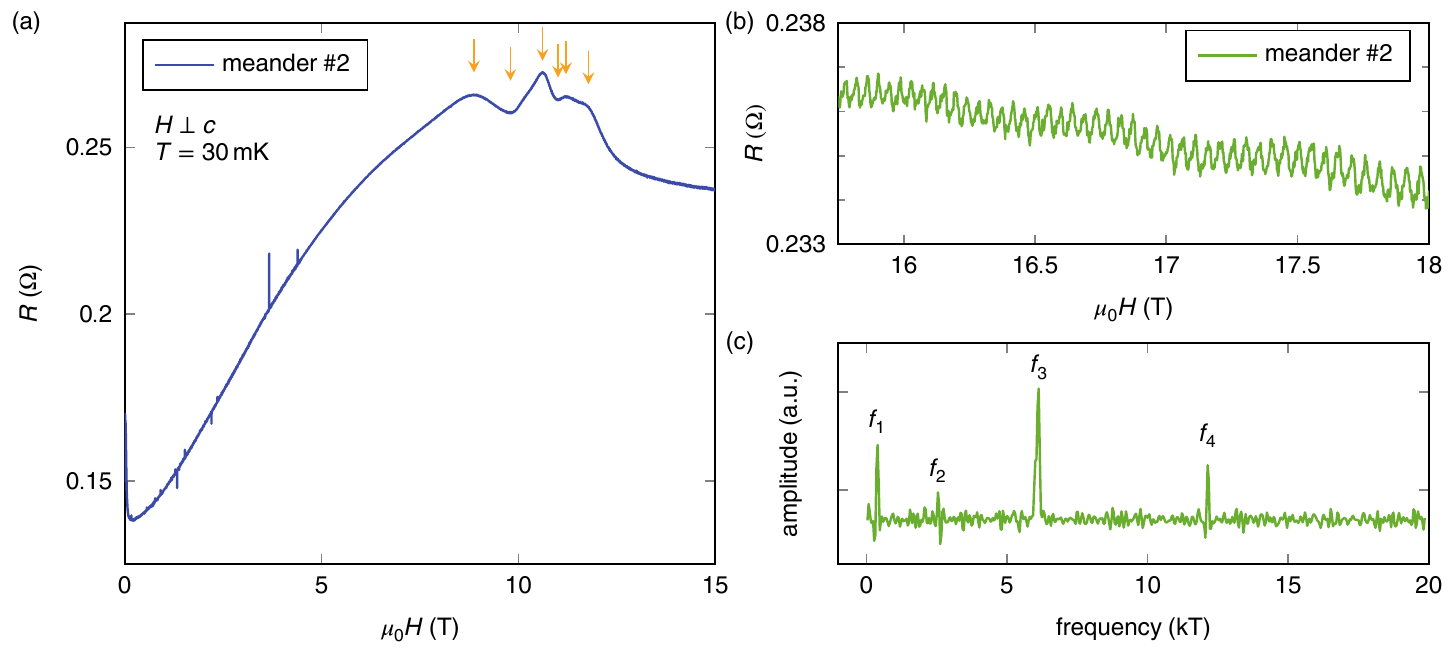}
	\caption{(a) Magnetoresistance of \YRS\ measured on meander \#2. There are several sharp slope changes which indicate Lifshitz transitions~\cite{Pfau2013,Pourret2013}. In the field range up to 15\,T at least six can be detected, indicated here by orange arrows. (b) SdH oscillations measured on meander \#2 at a temperature of $30$\,mK. (c) 
Fourier transform of the magnetoresistance signal in high-field range field range between 11\,T and 18\,T. In this range strongest observed frequencies are located at 0.39\,kT, 2.55\,kT, 6.12\,kT, and 12.15\,kT.}

	\label{fig:Lifshitz_and_oscillations}
\end{figure}
 
We performed magnetic field-dependent resistance measurements on a second meander structure cut from a high-quality crystal of RRR = 270 described in Tab.~\ref{tab.Samples_strain.effect} as meander \#2. The higher quality of this crystal was achieved by optimizing the temperature profile together with the crucible material in a new furnace \cite{Kliemt2020}. It turned out, that shortening the growth time by about one third in comparison to Ref. \cite{Krellner} and changing the crucible material to graphite or glassy carbon lead to less contamination of the melt from the crucible, which was an issue in Al$_2$O$_3$-based crucibles.
The corresponding magnetoresistance is shown in figure~\ref{fig:Lifshitz_and_oscillations}(a). With the increase in crystal quality and measurement resolution we are able to observe signatures of several Lifshitz transitions up to 13\,T at fields comparable with previous investigations~\cite{Pfau2013,Pourret2013}. This supports that microstructuring neither decreased the quality of the crystal, nor significantly influenced its intrinsic properties. At the same time though the relatively strong mechanical contact to the substrate material leads to a strain effect due to the differential thermal contraction as discussed below (see section \ref{straineffect}).

Further evidence of the quality of the meander \#2 is the presence of SdH oscillations at field above 10\,T shown in figure~\ref{fig:Lifshitz_and_oscillations}(b). The observed frequencies of 2.55\,kT, 6.12\,kT and 12.15\,kT (figure~\ref{fig:Lifshitz_and_oscillations}(c)) are in agreement with previously observed groups of frequencies in \YRS\ attributed to the "doughnut" or "jungle-gym" sheets of the Fermi surface~\cite{Zwicknagl2011,Rourke2008,Westerkamp2009,Sutton2010}. The $ab$-plane orientation was not independently measured, although the values as high as 12.15\,kT  - similar to previous observations up to $\approx 14$\,kT for $B \parallel [110]$~\cite{Westerkamp2009,Sutton2010,Friedemann2013} - suggest that the field in our sample was oriented close to the [110] direction. Furthermore, the positions of the kinks in magnetoresistance (orange arrows in figure~\ref{fig:Lifshitz_and_oscillations}) correspond well to those seen in Ref.~\cite{Pourret2013} for $B \parallel [110]$. We additionally observe a low-frequency oscillation at 0.39\,kT which has never been observed before. A similar frequency has been found experimentally only in the non-magnetic reference compound LuRh$_2$Si$_2$ and attributed to a beat frequency associated with the "jungle-gym" sheet of the Fermi surface~\cite{Friedemann2013}, but the fundamental frequencies are not observed here. Renormalized band structure calculations show that the relevant states are minority states forming the "jungle-gym" Fermi surface sheet. They are strongly hybridized and are responsible for the topological transitions between 9 and 13\,T in which, eventually, small portions of the Fermi surface disappear~\cite{Pfau2013,Pourret2013}. So, the appearance of this new frequency at high fields corroborates the significant change of the topology of the Fermi surface and the difference to the low-field Kondo-lattice state~\cite{Kummer2015}.

\section{High-temperature noise spectroscopy on meander \#3}
The control of the sample geometry and therefore its absolute resistance allows to observe not only the resistance in greater detail, but also resistance {\it fluctuations}.
These fluctuations -- measured as the noise power spectral density (PSD) $S_R$ -- reveal valuable information about the low-frequency charge carrier dynamics, which usually remains hidden in the averaged DC resistance $R$. In a voltage measurement $V = R I$ for a given current $I$, the PSD is defined as %$S_V = \langle |\tilde{V}(\omega)|^2 \rangle$, 
\begin{equation}
S_V(f) = 2 \lim_{T_i \rightarrow \infty} \frac{1}{T_i} \left| \int_{-T_i/2}^{T_i/2} \delta V(t) e^{-{\rm i} 2 \pi f t} {\rm d}t \right|^2 % \equiv  \frac{2|\mathcal{F}[V](f)|^2}{T}, 
\label{eq:noise-def-1}
\end{equation}
where $\delta V(t) = V(t) - \langle V(t) \rangle_t $ denotes the deviation of the signal from its long-term mean value and $T_i$ the time interval of the measurement. Note that the time-averaged mean $\langle V(t) \rangle_t$ may be considered zero. Indeed, in the experiment, this ’DC-offset’ often is sought to be suppressed in a five-terminal experimental setup so that only the fluctuating part of the signal is amplified and analyzed, see below. 
Averaging over the measuring time $T_i$ determines the bandwidth of the measurement, where the low-frequency limit depends on the stability of the system and the patience of the experimenter.) It is $S_R = S_V/I^2$, and very often a $1/f$-type noise $S_R \propto 1/f^\alpha$ with a frequency exponent $\alpha \approx 1$ is observed \cite{Mueller2018}.
The {\it Wiener-Khintchine theorem} connects the autocorrelation function of the system, {\it i.e.} $\Psi(\tau) = \langle \delta V(t) \cdot \delta V(t + \tau) \rangle$, which characterizes the kinetics of the fluctuations and describes the charge carrier dynamics of condensed-matter system on the microscopic level, to the noise PSD: 
\begin{equation}
S_V(f) = 4 \int_0^\infty \Psi(\tau) \cos{(2 \pi f \tau){\rm d}\tau}.
\end{equation}
By inverting the Fourier transform, the autocorrelation function $\Psi(\tau)$ can, in principle, be deduced from measuring $S_V(f)$. For zero delay ($\tau = 0$) the autocorrelation function equals the variance of the signal $\sigma ^2 \equiv \Psi(0) = \int_0^\infty S_V(f) {\rm d}f  = \langle \delta V(t)^2 \rangle$.
Furthermore -- unlike the linear (ohmic) part of the DC resistance, which measures the second moment of the current distribution in a sample -- resistance fluctuations measure the fourth moment and hence are highly
%the resistance fluctuations are 
sensitive to local variations of electric fields and the current density \cite{Kogan1996}. %, see section\,\ref{non-linear}.

Besides measurements of the thermal (Johnson) noise, which --- by exploiting the Nyquist theorem --- is an elegant way to determine the absolute resistance without driving a current through the sample, see section \ref{ultra-low temperature noise} below, structuring of \YRS\ on the micrometer scale allows to detect another type of noise, namely the intrinsic $1/f$-type fluctuations of a sample. Whereas thermal noise is frequency independent (‘white’) and therefore entirely uncorrelated, the $1/f^\alpha$ noise with a frequency exponent $\alpha$ close to 1 (in most cases between 0.8 and 1.4) reflects the kinetics of the charge fluctuations and therefore enables to access the microscopic motion and transitions of particles and their coupling to magnetic, structural or other degrees of freedom \cite{Kogan1996}. A power spectral density (PSD) $S \propto 1/f$ is ubiquitous in nature and is found in many condensed-matter systems like semiconductors and devices, metals and insulators, granular systems, magnetic thin films and sensors, tunnel junctions, spin glasses, superconductors, colossal magnetoresistance materials etc, reflecting the low-frequency dynamics of charge carriers \cite{Raquet2001,Mueller2011,Mueller2018}. 

In correlated electron systems, percolative superconductivity in the coexistence region of antiferromagnetic insulating and superconducting phases \cite{Mueller2009}, the critical slowing down of the charge carrier dynamics at the critical endpoint of a Mott transition \cite{Hartmann2015}, or the formation of a newly-discovered charge-glass state \cite{Sasaki2017,Thomas2022} have been investigated using $1/f$-noise spectroscopy, to name only a few examples. In the wide class of HF compounds, however, to the best of our knowledge so far no systematic measurements of this kind have been performed. For example, possible fluctuations coupled to the heavy quasiparticles upon entering the superconducting phase or at the Kondo temperature, and measurements of the intrinsic fluctuations in the vicinity of a quantum critical point seem attractive to explore. To a large part, the lack of such experiments in high-quality samples is due to the general difficulty to observe the intrinsic resistance fluctuations in highly-conductive systems, since the resistance noise PSD, $S_R$, scales with $R^2$, i.e.\ for samples with a low absolute resistance $R$, the noise PSD $S_R$ may be below the noise floor of the experimental setup. Considering a sample with active length $\ell$ and cross section $A$ and Hooge’s empirical law \cite{Hooge1969}
\begin{equation}
    S_R = \gamma_H \frac{R^2}{n_c \Omega f} = \gamma_H \frac{\rho^2}{n_c f} \cdot \frac{\ell}{A^3},
    \label{Hooge}
\end{equation}
where $\gamma_H$ denotes a material parameter characterizing the magnitude of the fluctuations of a system, $n_c$ the charge carrier density, $\Omega$ the sample volume under investigation, and the resistivity $\rho$. Clearly, the geometry factor $\ell/A^3$, and therefore the signal to be measured, can be easily enhanced by a factor 10 to 100 in a meander structure as compared to a regular-shaped sample, thereby making $1/f$-noise spectroscopy feasible in materials where it would not be for bulk crystals.

\begin{figure}[ht!b]
    \centering
    \includegraphics[width=\linewidth]{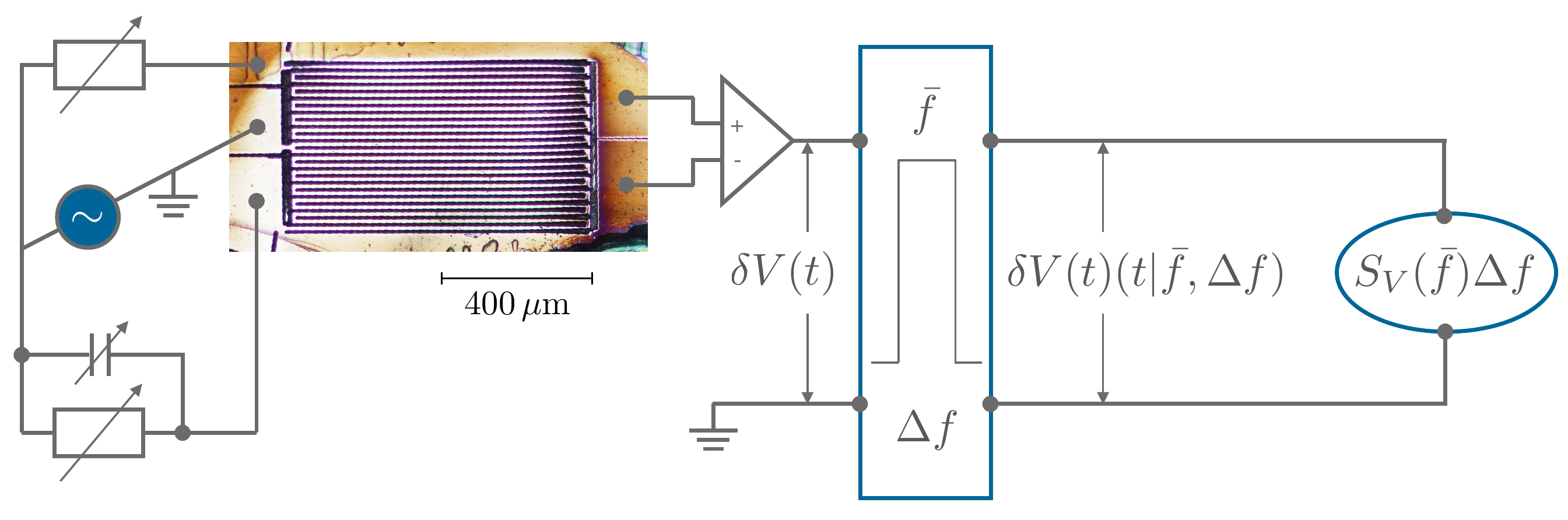}
    \caption{Schematic of the experimental five-point setup for low-level noise measurements on meander \#3. On the input side (left) the balance circuit of the Wheatstone bridge is depicted with the middle contact of the sample’s meander structure providing almost perfect symmetry of the two arms being part of the bridge. On the output side, only the fluctuating part of the voltage signal is first amplified, then processed by a band-pass adjustable-frequency filter with narrow bandwidth and an output detector that responds to the mean square of the signal (after \cite{Kogan1996}). In practice, noise sidebands produced by the resistance fluctuations that modulate the sinusoidally excited carriers are demodulated by a phase-sensitive detector (lock-in amplifier). The output signal of the lock-in amplifier is then processed by a spectrum analyzer which calculates the spectral density of the voltage fluctuations \cite{Scofield,Kogan1996,Mueller2011}.}
    \label{fig:Noise1}
\end{figure}

\begin{figure}[ht!b]
    \centering
    \includegraphics[width=\linewidth]{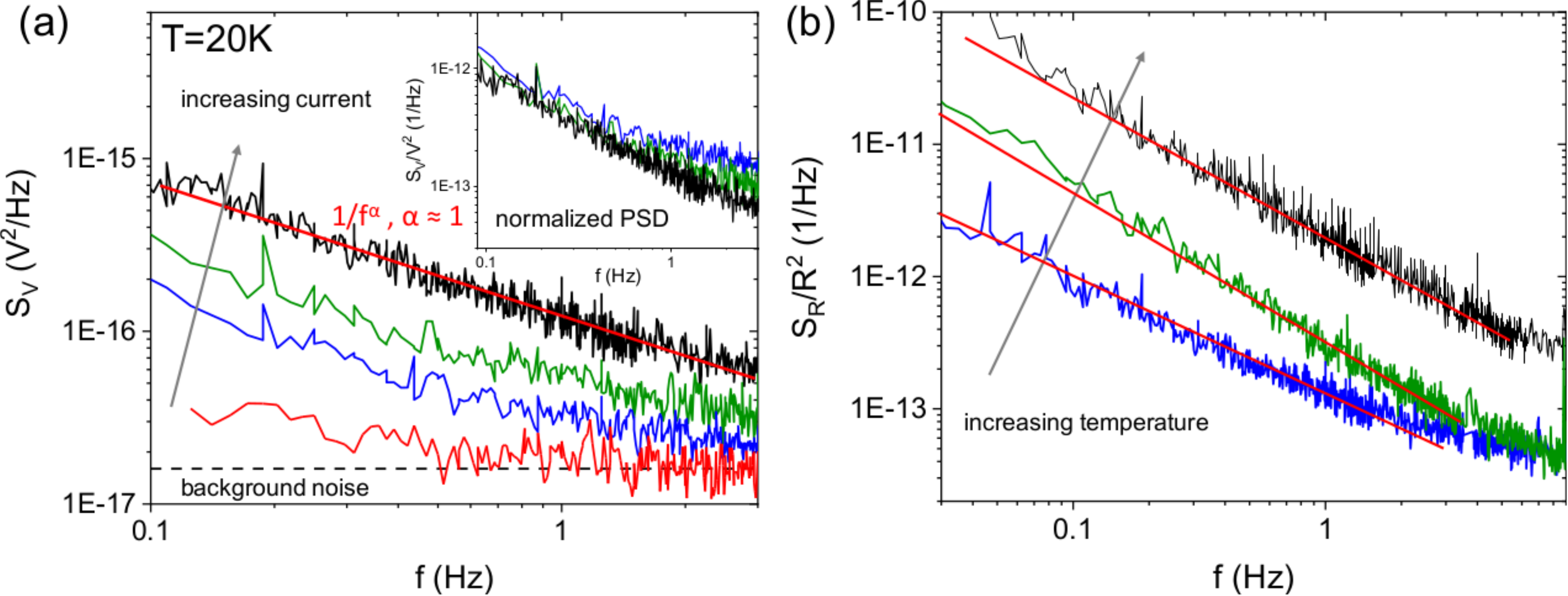}
    \caption{(a) Voltage noise PSD $S_V(f)$ at $T=20$\,K for increasing currents from $I = 175\,{\rm \upmu A}$ (red) up to $I = 880\,{\rm  \upmu A}$ (black) in a log-log representation showing an almost ‘white' noise for low currents and a $1/f$ behavior for higher currents which scales with the square of the current as shown in the inset (see text). (b) Normalized PSD $S_R/R^2$ at three different temperatures ($T = 20$\,K, 75\,K, 110\,K) demonstrating changes both in the noise magnitude and the frequency exponent (slope).}
    \label{fig:Noise2}
\end{figure}

Besides enhancing the absolute resistance, the microstructuring of a sample allows to make use of a particular electronic circuit with which in an AC (lock-in) measurement the sample’s resistance fluctuations can be shifted above the noise floor, often determined by the preamplifier \cite{Scofield}. Figure \ref{fig:Noise1} depicts such a five-terminal setup, where the sample itself (meander \#3, from same growth as meander \#2) forms part of a Wheatstone bridge, thereby allowing to balance the DC offset such that only the fluctuating part of the signal is amplified and any perturbations as, e.g., thermoelectric effects, fluctuations of the voltage or current source and those of the thermal bath temperature are efficiently suppressed \cite{Scofield}.  
%[Scofield, J.H. AC method for measuring low-frequency resistance fluctuation spectra. Rev. Sci. Instrum. 1987, 58, 985–993].	

Figure \ref{fig:Noise2}(a) shows the voltage noise PSD of a microstructured meander sample \YRS\  
at $T = 20$\,K for various applied currents measured in the above described five-terminal setup using a homemade resistance bridge, a voltage preamplifier SR554 and a lock-in amplifier SR830. Shown is the averaged voltage noise PSD calculated by a signal analyzer SR785, where typically 30 - 50 averages are sufficient. 
For low currents, a nearly frequency-independent behavior determined by the noise floor of the electronic setup is found. Clearly, $S_V$ increases with increasing current and for larger currents a distinct $1/f$-type behavior $S_V \propto 1/f^\alpha$ with $\alpha \approx 1$ is observed. The inset shows the expected scaling $S_V \propto V^2 \propto I^2$ which is a main condition necessary to prove that $1/f$-type fluctuations intrinsic to the sample resistance are observed, besides other checks to exclude the influence of spurious noise sources \cite{Mueller2011}. 
Figure \ref{fig:Noise2}(b) shows the normalized resistance noise $S_R/R^2 = S_V /(I^2\cdot R^2)$ for different temperatures demonstrating that temperature-dependent measurements showing variations of the intrinsic $1/f$-type fluctuations, which reflect changes in the low-frequency charge carrier dynamics, are feasible in metallic HF compounds with low impedances. For the present sample, we estimate a Hooge parameter $\gamma_H$ in Eq.\ (\ref{Hooge}) of order $10^5$ using an estimated sample volume of $\Omega = 4.68\cdot10^{-12}\, {\rm m}^{-3}$ and a charge carrier density $n_c=2.6 \cdot10^{28} {\rm m}^{-3}$, see Ref.\cite{Paschen}. The relative $1/f$-noise level of \YRS\ estimated in this way is quite large compared to, e.g., clean and homogeneous semiconductors with typical values $\gamma_H \sim 10^{-3} - 10^{-2}$. However, $\gamma_H$ in different materials has been found to vary over many orders of magnitude from $10^{-6}$ to $10^7$~\cite{Raquet2001}. It should be noted, however, that the assumption of $1/f$ -noise being composed of independent fluctuation processes of individual mobile charge carriers in many cases is a crude oversimplification, see Ref.\cite{Mueller2018} and references therein. Therefore, the physical meaning of $\gamma_H$ as introduced above is limited, in particular since the noise magnitude usually is strongly temperature dependent. However, it is still useful to compare the relative noise level of different compounds or different samples of the same compound via the parameter $\gamma_H$, which may serve as an additional parameter to characterize a sample. Furthermore, the temperature dependence of the noise magnitude $S_R/R^2(f)$, often conveniently determined at $f = 1$\,Hz, and the frequency exponent $\alpha = - \partial \ln{S_R/R^2}/ \partial \ln{f}$ provide information on the distribution and shift of spectral weight as a function of temperature of the low-frequency (collective) excitations. An analysis of the temperature dependence of $\gamma_H$ and $\alpha$ will be published elsewhere. In this work we demonstrate that such measurements are feasible for high-quality low-impedance HF compounds. 

\section{Tuning of $T_N$ by strain on meander \#2 and bar \#1\label{straineffect}}
\begin{figure}[t]
	\centering
	\includegraphics[width=1\linewidth]{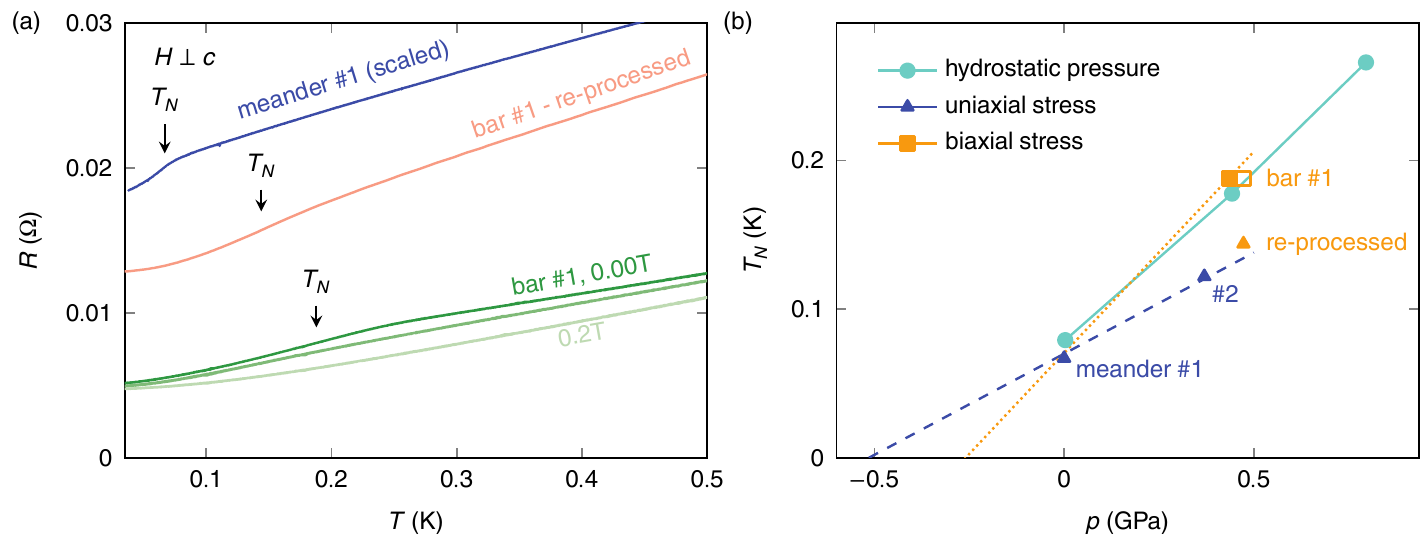}
	\caption{(a) Low-temperature resistance measurements of meander and bar samples. The difference in thermal contraction between substrate and samples leads to a shift of $T_N$ which is geometry-dependent. Whereas in meander \#1 no shift occurred, the transition into the AFM state in the bar-shaped sample onsets at about 188\,mK. By partial removal of epoxy below the sample (bar \#1 re-processed) this shift can be reversed. 
	(b) Comparison of N\'eel transition temperatures in \YRS\ for hydrostatic pressure~\cite{Mederle2002} and uni- or biaxial stress due to differential thermal contraction in structured samples. The stress of the samples is estimated using the Ehrenfest relation and is predominantly uniaxial (dashed) in the meander samples and biaxial (dotted) in the bar \#1 sample (closed symbols). Alternatively, the stress can be estimated from the differential thermal contraction (open symbol). With re-processing the stress character can be modified in bar \#1 from biaxial to uniaxial.}
	\label{fig:TNshift2}
\end{figure}

\marginnote{Strain effect}
Control over sample geometry opens up further possibilities besides the improvement in the observed signal or its fluctuations. One of these is tuning of interactions through the application of strain from a substrate, either uni- or biaxially. This has been shown to influence the superconducting transition temperature in CeIrIn$_5$\cite{Bachmann2019} and can in general be used to shift phase transitions of hard \cite{Park2020} and soft \cite{Bartlett2021} correlated electron systems. 
With \YRS\ being located in the close vicinity of a QCP we report here on the first attempts to use this parameter to influence its N\'eel transition temperature. We expect that any positive uniaxial pressure independent of its direction will increase $T_{N}$ because the discontinuities in the thermal expansion coefficient at $T_N$ of \YRS\ exhibit positive jumps for both the basal plane ($\perp$) and the $c$-axis ($\parallel$)~\cite{Kuechler2005}. With their magnitudes $\Delta\alpha_{\perp} = (9.8\pm 0.5) \times 10^{-6}$\,K$^{-1}$, $\Delta\alpha_{\parallel} = (6.5\pm 0.5) \times 10^{-6}$\,K$^{-1}$, we use an Ehrenfest relation
\begin{equation}\label{Ehrenfest1}
    2\frac{dT_{N}}{dp_{\perp}} + \frac{dT_{N}}{dp_{\parallel}} = \frac{dT_{N}}{dp} = \frac{T_{N}V_{\mathrm{mol}}}{\Delta C}\left( 2\Delta\alpha_{\perp} + \Delta\alpha_{\parallel} \right)
\end{equation}
to estimate the response to pressure along the different directions. Here, with $\Delta C \approx 0.24$\,J/mol$\,$K~\cite{Custers2003}, the pressure dependence of $T_{N}$ at $p = 0$ is $dT_{N}/dp \approx 0.36\,$K/GPa which slightly overestimates the response observed in hydrostatic pressure experiments~\cite{Mederle2002} (see figure~\ref{fig:TNshift2}(b)). 

We observe indeed a significant shift of $T_N$ in meander \#2 which was used for the experiments shown in figures~\ref{fig:Lifshitz_and_oscillations} and \ref{fig:YRS_sc}. It was cut from a high-quality sample with an RRR of 270 which grew in the shape of a thin platelet. This platelet was then glued to a silver sample holder using a thin layer of Stycast 1266 epoxy and cured at \SI{60}{\celsius} (see Tab.~\ref{tab.Samples_strain.effect}). The relatively stiff epoxy (Young's modulus of 11 GPa at 77 K~\cite{PereaSolano2004}), applied in a thin layer with a thickness of less than $100\,\mathrm{\upmu}$m leads to a significant transmission of the differential thermal contraction between the sample and the silver sample holder. This strain - which in this sample geometry is primarily uniaxial - leads to an increase of $T_N$ from 67\,mK to 120\,mK in meander~\#2 (figure \ref{fig:TNshift2}(b) and \ref{fig:YRS_sc}(b)). Such a shift of $T_{N}$ was not found in the meander \#1 which was attached to the substrate with a softer epoxy (Araldite Rapid). 

We now estimate the magnitude of the shift in comparison with the differential thermal contraction through the substrate. As an estimate for the transmitted strain we can consider the thermal contraction of the silver substrate from room temperature to cryogenic temperatures of $\Delta L(T\rightarrow 0)/L \approx -0.4\%$\cite{Pobell2007}. For \YRS\ no complete data of thermal expansion from room temperature to low temperatures are available. From powder X-ray diffraction measurements down to $40\,$K the contraction in the $ab$-plane $\Delta a(T\rightarrow 0)/a$ is at least $-0.08\%$.  An alternative estimate can also be obtained 
by averaging values among similar systems (CeAl$_3$, YbNi$_2$B$_2$C, YbAlB$_4$) found in literature. The mean of their volume thermal expansion coefficient $\beta$ between 0 and 300\,K is $5 \times 10^{-6}$\,K$^{-1}$ ($\Delta L(T\rightarrow 0)/L \approx -0.15\%$)~\cite{Kagayama1990,Schmiedeshoff2009,Matsumoto2017}. 

This results in a net compressive strain $\epsilon_{0}$ between $-0.25$\% and $-0.32$\%  for a thin microstructured sample of \YRS\ attached with stiff epoxy to a silver substrate.
In practice the strain transmission with such a meander-shaped sample geometry is limited by details of the interface epoxy between the sample and the substrate. Especially for thicker epoxy layers the transmitted strain decreases substantially and we expect that only thin samples approach this upper bound.
\begin{table}[t]
\centering
\caption{Overview over various properties of the microstructured samples. Meander \#1 is from an earlier growth using Al$_2$O$_3$ crucibles, meander \#2, and bar \#1 are from an optimized growth approach \cite{Kliemt2020}. }
\label{tab.Samples_strain.effect}
\begin{tabular}{ccccc}
\br
%Sample          & FIB      		& YRK-2AA   & YRK-2AE	& YRK-2AE      \\
Sample			 & meander \#1   	& meander \#2& bar \#1		& bar \#1 re-processed  \\
\mr
%measurement	 & R03				$ R06		& R07		& R09\\
RRR 			 & 35 				& 270		& 180		& 180\\
Epoxy used        & Araldite Rapid	& Stycast 1266 	& Stycast 2850 	& partially removed\\
Thickness        & 15-75$\,\upmu$m  	& 10$\,\upmu$m& 10$\,\upmu$m& 10$\,\upmu$m        \\
Shape 			 & meander  		& meander   & bar      & bar  \\
$T_N$        	 & 67$\,$mK 		& 120$\,$mK & 188$\,$mK & 144$\,$mK            \\
%$H_c$            & 65$\,$mT 		& 80$\,$mT  & 107$\,$mT & ?$\,$mT              \\
Image     &\includegraphics[width=0.19\textwidth]{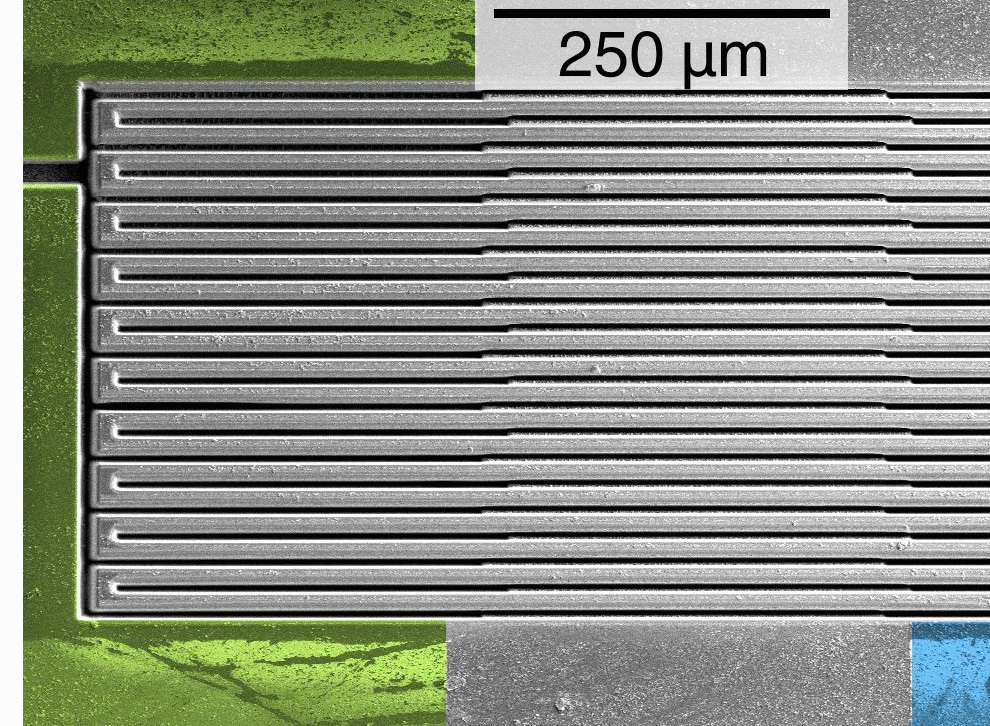}&\includegraphics[width=0.19\textwidth]{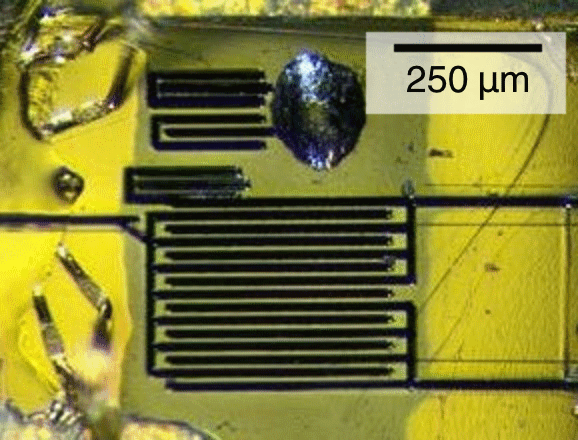} & \includegraphics[width=0.19\textwidth]{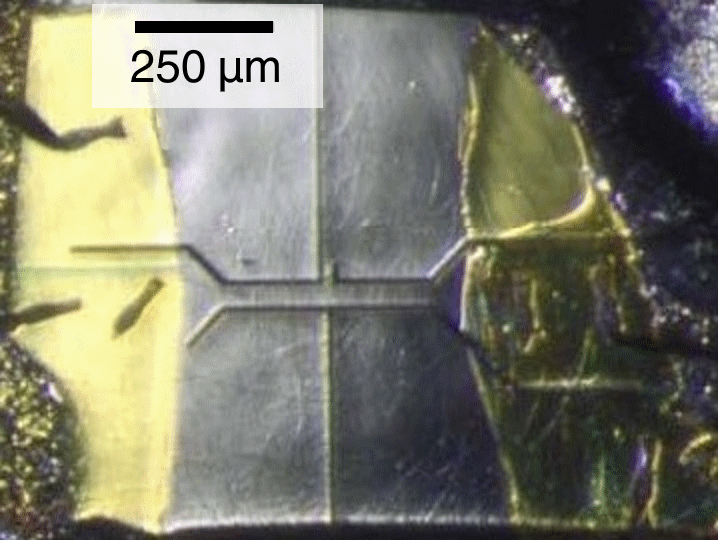} & \includegraphics[width=0.19\linewidth]{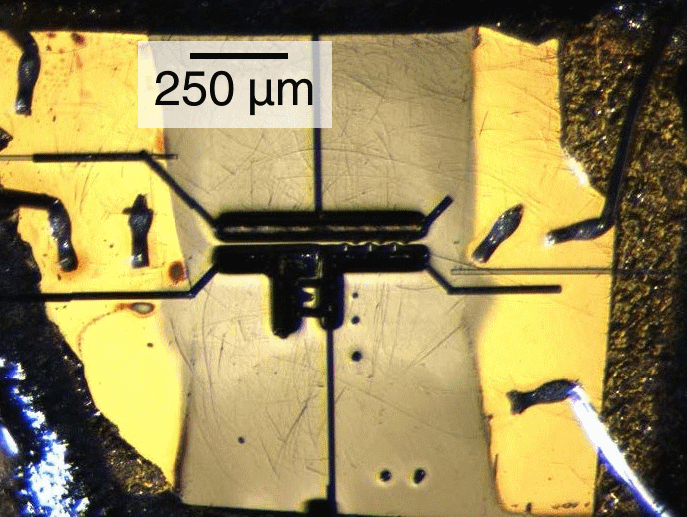} \\
\br
\end{tabular}
\end{table}

One technique to increase the transmitted strain is a sample geometry consisting of a long bar with large taps at each side (see Tab.~\ref{tab.Samples_strain.effect}). In such an optimized geometry a significant amount of force is transmitted through the taps at the side with much larger contact area to the substrate than through the bottom of the bar alone~\cite{Park2020}. Additionally we used a filled epoxy, Stycast 2850, which further strengthens the mechanical interface between sample and substrate. Here the increase of the N\'eel temperature is highest in this sample with $T_N \approx 188$\,mK (see figure~\ref{fig:TNshift2}). To distinguish the strain contribution from the interface at the bottom of the bar and from the side taps we performed another FIB procedure and removed the Stycast layer underneath the bar up to the taps. After this step $T_N$ is reduced to \SI{144}{\milli\kelvin}, reducing the strain transmission to the underside of the bar, which supports the dominant transmission via the outer taps in accordance with finite element simulations of such geometries~\cite{Park2020}. A summary of parameters of the four microstructured samples is given in Tab.~\ref{tab.Samples_strain.effect}.

Motivated by the significant change of $T_N$ to higher temperatures, we can estimate which stress would be required to drive the transition temperature to zero. From equation~\eref{Ehrenfest1} we obtain for the in-plane strain component a slope $dT_{N}/dp_{\perp}$ of $0.15$\,K/GPa as shown in figure~\ref{fig:TNshift2}(b). To completely suppress the antiferromagnetic order we therefore require an in-plane tensile stress between -0.5 GPa and -0.25 GPa in, respectively, the uniaxial or biaxial case, as shown in the extrapolation in figure~\ref{fig:TNshift2}(b).

The corresponding strain can be estimated from measurements of the bulk modulus $B$ in \YRS\ if we disregard any anisotropy of the mechanical properties. Under these assumptions and using a Poisson's ratio of $\nu \approx 1/3$ the Young's modulus $E = 3B(1-2\nu)$ equals the bulk modulus. With a measured room temperature value of $B=189$\,GPa \cite{Plessel2003} the corresponding uniaxial tensile strain of $\epsilon_{\perp} = -dp_{\perp}/E = -0.26\,\%$ is rather high and possibly beyond the elastic limit of the material. On the other hand a mismatch in thermal expansion between sample and substrate leads to a biaxial stress and, under ideal circumstances, the required strain is reduced into two equal uniaxial components of $\epsilon_\mathrm{a,b} \approx -0.13\,\%$.  

\section{Ultra-low temperature noise measurements on meander \#2}
\label{ultra-low temperature noise}
\begin{figure}[t]
	\centering
	\includegraphics[width=\linewidth]{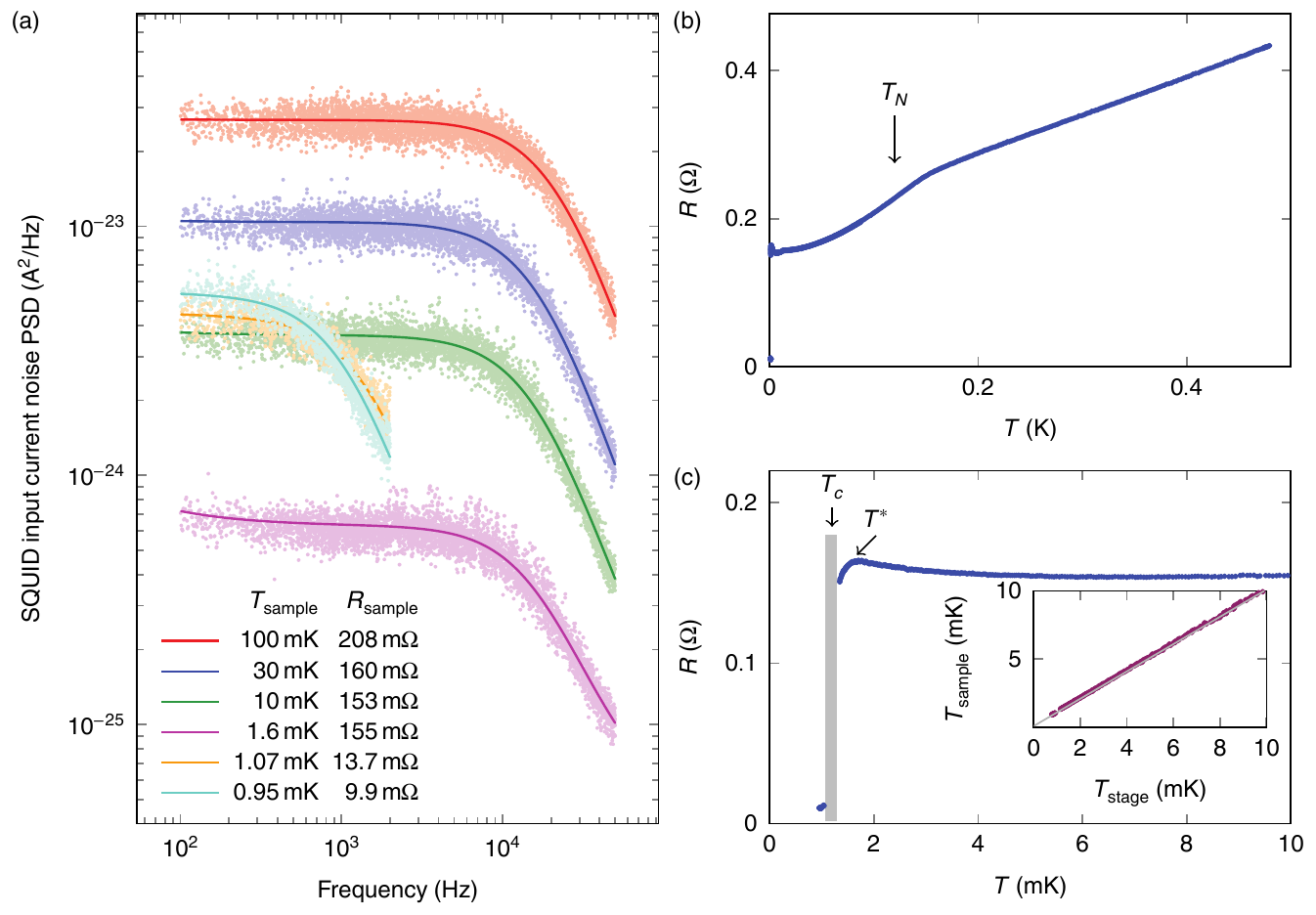} 
	\caption{Noise measurements of meander \#2 down to ultra-low temperatures at Earth's field: (a) Power spectral density (PSD) of the current measured by the SQUID current sensor as a function of frequency and fits according to Eq.~(\ref{eq:CSNT}). (b) Temperature dependence of the resistance of the meander extracted from noise measurements across the AFM transition at $T_N = 120$\,mK. The transition is as sharp as that measured in bulk samples (see figure~\ref{fig:YRS_fig2_rho_n_H}). (c) Resistance of the meander for temperatures below 10\,mK. The lowest sample temperature was 0.95\,mK. The grey region indicates the temperature range $T_c = 1.2 \pm 0.1$\,mK, were strong resistance fluctuations prevent reliable estimation of the sample resistance and temperature. We attribute $T_c$ with the superconducting transition, as the resistance drops by more than an order of magnitude across this narrow temperature range, with an onset at $T^* = 1.6$\,mK. The inset shows a linear plot of the sample temperature versus the stage temperature, indicating excellent thermalization of the meander down to 1\,mK.}
	\label{fig:YRS_sc}
\end{figure}
When designing an ultra-low temperature experiment, one has to consider the lowest temperature $T_{\mathrm{s}}$ that a metallic sample can reach, if one point of the sample is well-thermalized to the refrigerator at temperature $T_{0}$. Electron-phonon coupling is weak at ultra-low temperatures,
and typically the cooling is dominated by electrons obeying the Wiedemann-Franz law in the cooling contact and the sample, giving 
\begin{equation}\label{eq:Ts:vs:Q}
T_{\mathrm{s}}^2 - T_{0}^2 \approx (P_{\mathrm{diss}} + \dot{Q}_{0})(R_{\mathrm{s}} + R_{\mathrm{c}}) / L_{0},
\end{equation}
where in addition to the dissipation $P_{\mathrm{diss}}$ in the resistance measurement introduced above, we consider the spurious heat leak to the experiment $\dot Q_{0}$. Here $T_{0}$ is the temperature of the refrigerator
and $L_{0} = 2.44\times10^{-8}$\,W$\Omega$/K$^{2}$ is the Lorenz number.
Details such as the position of the cooling contact within the sample and whether $\dot Q_{0}$ is generated uniformly within the sample, lead to prefactors of order unity in Eq.~(\ref{eq:Ts:vs:Q}).
For the nuclear demagnetisation refrigerator with base temperature $T_{0} = 0.5$\,mK, $R_{\mathrm{s}} + R_{\mathrm{c}} = 0.1\,\Omega$, the applied power must be $P_{\mathrm{diss}} + \dot Q_{0} < 0.2$\,pW in order to reach $T_{\mathrm{s}} < 1$\,mK.

The meander \#2 is well-suited to ultra-low temperature experiments due to its low contact resistance, achieved by evaporating a gold film on top of the contact pads.
Furthermore the pad in the middle of the meander was chosen for cooling, and this was connected to the electrical and thermal ground via two ultrasonically bonded 25\,$\upmu$m Au wires in parallel.
At 4\,K we measured 3.54\,$\Omega$ across the whole meander, while each half to ground was 1.56\,$\Omega$ and 1.98\,$\Omega$.
Since these numbers add up, the 10\,m$\Omega$ uncertainty of these measurements gives the upper bound on contact resistance $R_{\mathrm{c}}$, satisfying $R_{\mathrm{c}} \ll R_{\mathrm{s}}$.

For the initial ultra-low-temperature investigation we configured this microstructure as a SQUID-based current-sensing noise thermometer~\cite{Casey2014}.
The outer contact pads of the meander were connected to Nb/NbTi leads of the SQUID current sensor with ultrasonically-bonded 25\,$\upmu$m Al wires. The experiment was enclosed in a Nb shield trapping Earth's magnetic field.
The chosen non-dissipative technique with no direct connections to room-temperature electronics ensured $P_{\mathrm{diss}} = 0$ and $\dot Q_{0} < 1$\,pW~~\cite{Casey2014}.
The sample temperature $T_{\mathrm{s}}$ and the resistance $R$, that includes the meander and the contacts to Al, were inferred from fitting the measured power spectral density (PSD) of the SQUID input current noise, figure~\ref{fig:YRS_sc}(a), to
\begin{equation}\label{eq:CSNT}
    S_{I}(f) = \frac{4k_{B} T_{\mathrm{s}} R}{R^{2} + 4\pi^2 f^{2}L^{2}} + N + N'/f,
\end{equation}
where $L = 1.51$\,$\upmu$H is the input inductance of the SQUID and $N$ and $N'$ characterise fixed background noise and are determined from noise above 100\,kHz and below 100\,Hz. Here $N = 4.47 \times 10^{-26}$\,A$^2$/Hz and $N' = 9.19 \times 10^{-24}$\,A$^2$.

The fit results, figure~\ref{fig:YRS_sc}(b), demonstrate a N\'eel transition at $T_{N}$ between linear temperature dependence of the resistance in the non-Fermi liquid state and quadratic in the antiferromagnetic Fermi-liquid. By comparison with conventional resistance measurements we conclude that $R$ is dominated by the resistance of the meander.
On cooling below 10\,mK, figure~\ref{fig:YRS_sc}(c), we observe a weak increase in the resistance just before the steep decrease with onset at $T^* = 1.6$\,mK.
Between 1.1 and $1.3$\,mK (grey area in figure~\ref{fig:YRS_sc}(c)), strong resistance fluctuations were observed, and the sample resistance and temperature could not be reliably inferred.
The resistance drops by more than an order of magnitude over this temperature range,
therefore we associate it with the superconducting transition temperature $T_{c} = 1.2 \pm 0.1$\,mK.
The residual $R = 10$\,m$\Omega$ below $T_{c}$ is consistent with the U-sections of the meander, which experience different strain than the rest of the microstructure, remaining normal.

We observe that $T^{*}$ is close to the temperature $T_{A} = 1.5$\,mK of the electro-nuclear magnetic transition~\cite{Schuberth2016, Knapp2022}, and suggests that $T^{*}$
may be the resistive signature of $T_{A}$. In this scenario, the superconductivity only emerges in the uniformly-strained structure below $T_{A}$, unlike the transport signatures reported in bulk crystals at 6-8\,mK~\cite{Nguyen2021,Levitin2022}.
Alternatively, $T_{A}$ may coincide with $T_{c}$, as suggested in Ref.~\cite{Schuberth2016}, both being shifted down due to the strain that increases $T_{N}$ with $T^{*}$ being the onset of $T_{c}$ and not a separate feature.

The sample temperature was found to be in good agreement with the temperature of the refrigerator, reaching $T_{\mathrm{s}} = 0.96$\,mK at $T_0 = 0.79$\,mK.
Estimating the heat leak $\dot Q_{0}$ is more straightforward in the normal state, where no strong violations of the Wiedemann-Franz law are expected. Using the observed $T_{\mathrm{s}} = 1.60$\,mK at $T_{0} = 1.44$\,mK
we estimate $\dot Q_0 \approx 0.1$\,pW from Eq.~(\ref{eq:Ts:vs:Q}).
Thorough shielding of the sample and filtering of the measurement lines~\cite{levitin2022cooling} are expected to reduce $\dot Q_0$ by 1-2 orders of magnitude, opening temperatures well below 1\,mK to microstructured devices and enabling driven resistance measurements, that can clearly distinguish between the sample and contact resistances.

\section{Conclusion}
To extract the intrinsic properties of high-quality materials by low-temperature resistance measurements remains challenging due to the trade-off between dissipation and obtained signal amplitude. With FIB microstructuring we demonstrated an increase in resistance in samples of \YRS\  by several orders of magnitude without negatively altering their characteristics and physical properties. The sharpness of the AFM transition in resistivity at $T_{N} \approx 70$\,mK  and the onset of superconductivity below $T_{c} \approx 1.2$\,mK, expected only for high-quality and homogeneous samples, provide strong evidence for this. Moreover, the larger resistance provides a significant increase in resolution and therefore revealed quantum oscillations signatures in resistance in fields between 13 and 18\,T. 
This process enables noise spectroscopy measurements of metallic samples, which requires that the contact resistance becomes negligible compared to the sample resistance. Here we observe typical $1/f$-type fluctuations in high-temperature noise spectra in a microstructured sample. By going to ultra-low temperatures, SQUID-based noise measurements demonstrate that a microstructured sample can be cooled down below 1\,mK while measuring simultaneously its intrinsic resistance and temperature. 
Finally, with thin and structured samples we could induce a significant shift of the N\'eel transition to up to 188\,mK from the differential thermal expansion between samples and substrates through a thin epoxy layer. This allows for different material combinations to apply both compressive and tensile stress, with the potential to reach a stress-induced QCP in \YRS. 
Going beyond the two-dimensional structures shown here allows detailed analysis of anisotropy effects, but also requires a more complex fabrication process which will be part of a future study~\cite{Shirer2022}.

\ack
The authors would like to thank 
A. Casey, S. Friedemann, M. Garst, P. Gegenwart, C. Geibel, R. K\"uchler, S. Lausberg, P. J. W. Moll, Y. Ny\'eki, K. Semeniuk, K. R. Shirer, F. Steglich and G. Zwicknagl for motivating discussions. P. Wein and C. Butzke for sample preparation and evaluation of thermal expansion data. J. M. Ellinghaus for testing the micro-milling properties of \YRS\ (Fraunhofer IPT, Aachen, Germany) and V. Antonov (RHUL, London) for aluminium and gold wirebonds. We acknowledge funding by the German Research Foundation (DFG) via the TRR 288 (422213477, project A03, A10 and B02) and projects KR3831/4-1 and BR 4110/1-1. This work was supported by the EU H2020 European Microkelvin Platform EMP, Grant No. 824109.

\section{Appendix}
\subsection{Microstructuring possibilities}
We considered alternative microstructuring methods which including milling, electro discharge machining (EDM) or laser cutting which can be more flexible but are less suitable for this type of application. 

\marginnote{Milling}
Micromachining using milling of single crystalline materials is difficult because instead of shaving off chips from the workpiece large platelets break from the edges. In tests with \YRS\ using a $50\,\upmu$m diameter high speed steel end mill and slot widths between 100 to $200\,\upmu$m were achievable, although with significant damage from the milling procedure \cite{Ellinghaus}. In oxide materials such as LiNbO$_3$ channels with widths of $300\,\upmu$m were milled using a diamond-coated bit with relatively few surface defects \cite{Huo16}. This approach would need to be scaled to much smaller channel widths though to achieve large resistance changes.
\marginnote{Electro discharge machining}
Cutting millimeter sized bars from larger single crystals using EDM is an established technique for high purity materials, used for example to cut needles out of the unconventional superconductor UPt$_3$~\cite{Kycia98}. Depending on the metallurgy and machining parameters the surface damage can be minimized but the heat affected zone where significant changes to the crystalline properties occur extends tens of microns into the material~\cite{Bleys06}. 
It is therefore unsuitable for small slot widths or requires additional preparation steps such as chemical etching after the machining.
\marginnote{Laser cutting}
When using a laser to cut into a material along the edge of the cutting path melting leads to changes of the crystalline structure. The width of this heat affected zone is comparable to that obtained by using EDM. In experiments using chromium-molybdenum steel the material properties converge to bulk properties in a depth of approximately $50\,\upmu m$ from the cut surface \cite{Singh08}.

\subsection{FIB structuring technique}
In general, FIB-structuring can be used to pattern bulk materials into customized geometries which allow to directly access the sample properties under investigation. For example, FIB patterning has been used to prepare samples with high aspect ratios to investigate materials with very low resistivity by means of electrical transport~\cite{Moll2016} or to align the device geometry precisely with the crystalline axes to study orientation-dependent phenomena~\cite{Moll2014}. For any sample geometry, the detrimental effect on the sample material due to the exposure to ion beams can be kept at a minimum by applying an appropriate sequence of process steps, e.g. using grazing incidence in the final polishing steps. The damage to the sample such as implantation of Ga atoms and distortions to the lattice structure depends on material properties, e.g. composition, but is typically limited to a few 10 nm from the surface~\cite{Moll2018}.

\section{References}
\bibliography{literature}
\bibliographystyle{iopart-num.bst}

\end{document}